# Carbon Nanotube Fabrication at Industrial Scale: Opportunities and Challenges

Joydip Sengupta

*Department of Electronic Science, Jogesh Chandra Chaudhuri College, Kolkata, West Bengal, India*

## 9.1 INTRODUCTION

### 9.1.1 Importance of Nanomaterial

The discovery of new materials with unique properties is the principal parameter for the sustained development of contemporary devices and for upliftment of device performances. In the last decade, intensive research efforts were made to create a large number of novel materials, notably those belongs to the nanometer regime. The outcome of the prolific research are the structures with reduced dimension, viz. two-dimensional structure, one-dimensional structure, and zero-dimensional structure. As the size of a material reduces to nanometer-scale dimensions then the material in general become superior to its bulk counterpart for many applications owing to its higher surface-to-volume ratio, size-dependent properties, and its potential for downscaling of device size.

Among different elements, Carbon, placed at group 14 (IV A), has become one of the most important elements in the periodic table owing to its ability to form $sp^3$, $sp^2$, and sp hybrids which results in 3D (diamond and graphite), 2D (graphene), 1D [carbon nanotube (CNT)], and 0D (Fullerene) materials with a wide variety of physical and chemical properties (Fig. 9.1).

### 9.1.2 Carbon Nanotube

Among the carbon allotropes, CNT has become a center of attraction in the field of nanoscale research in modern science. Nanotubes are nearly one-dimensional structure due to their high length to diameter ratio. CNTs exhibit a unique combination of electronic, thermal, mechanical, and chemical properties [2–5], which promise a wide range of potential applications in key industrial sectors such as nanoelectronics [4], biotechnology [6], and thermal management [7]. From a historical perspective tubular carbon nanostructures were first observed Fig. 9.2(A) as early as 1952 by Radushkevich and Lukyanovich [8]. A few years later Oberlin et al. [9] published clear images Fig. 9.2(B) of hollow carbon fibers with nanometer-scale diameters using a vapor growth technique. However, it was not until nearly two decades later, when Iijima [10] reported the observation of CNTs in the journal of Nature that worldwide interest and excitement was generated. Iijima's work is certainly responsible for the flare-up of interest in CNT research in the scientific community which resulted in the rapid development of this field. Iijima clearly observed the multiwalled CNTs (MWCNTs) while studying the soot made from by-products obtained during the synthesis of fullerenes by the electric arc discharge method (Fig. 9.3).

Since then the field has advanced at a breathtaking pace that is reflected in the increasing number of publications along with many unexpected discoveries.

### 9.1.3 Chemical Vapor Deposition (CVD)

There are many methods by which CNTs can be produced, including but not limited to arc discharge, laser ablation, and chemical vapor deposition (CVD). These three synthesis methods of CNT can be classified into two main categories depending on the growth temperature. High temperature routes are the electric arc method and the laser ablation method, whereas medium temperature routes are based on CVD processes. The high temperature process involves sublimation of graphite in an inert atmosphere and condensing the resulting vapor under a high temperature gradient. The difference between the various processes is the method used for subliming graphite. An electric arc formed between two electrodes is used for sublimation of graphite in case of the arc discharge method. An ablation induced by a laser is used for sublimation of graphite in case of the laser ablation technique. Previously, high-temperature processes were used as the primary methods for synthesizing good quality CNTs. However, both methods have major disadvantages. Firstly, both methods require high purity graphite rods, consume ample amounts of energy, and yield is also low. So, these methods are not commercially viable to scale up for CNT production at an industrial level. Secondly, CNTs grown by high-temperature methods are in highly twisted forms, assorted with unwanted species of carbon and metal. Thus the grown nanotubes are hard to clean, manipulate, and accumulate for construction of CNT-based device architectures. Controlled production on substrates with preplanned CNT structures has not been possible by these vaporization methods. These necessitate the application of CVD process by which all the above mentioned drawbacks can be overcome. CVD permits abundance of hydrocarbons in several state (solid, liquid, or gas), facilitates the use of different substrates, and harvests CNT in a variety of forms (powder, films), also in different shapes (straight, bamboo-like, coiled). Even site selective growth of in situ metal filled CNT growth on patterned substrate is possible with CVD, which proves the versatility of the method. Thus the present work focuses on CVD synthesis since this method can be easily extended to industrial-scale fabrication.

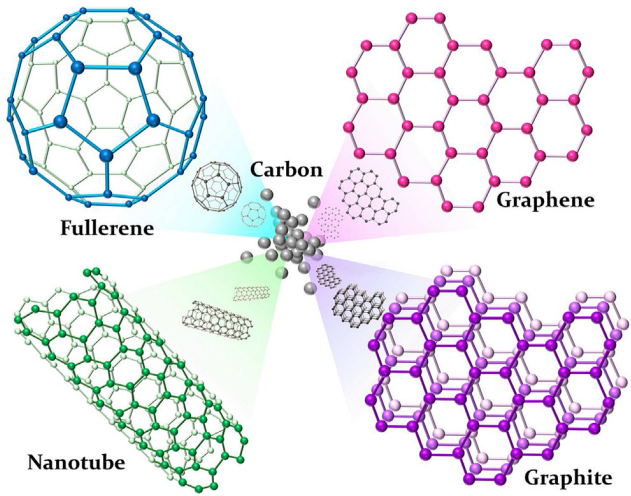

**FIGURE 9.1** Carbon allotropes in four different crystallographic structures [1].

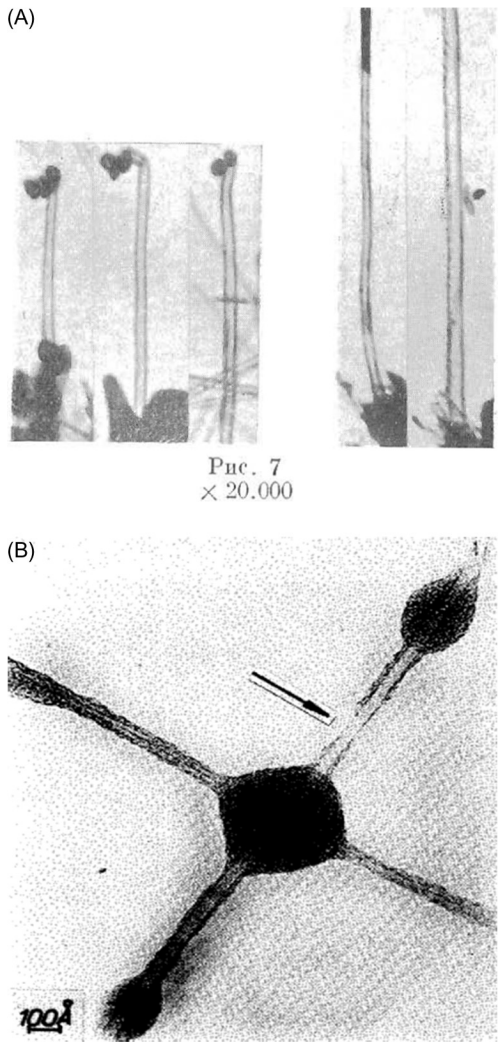

**FIGURE 9.2** Examples of transmission electron microscopy images of CNTs published by (A) Radushkevich et al. [8]; (B) Oberlin et al. [9]. *CNTs*, carbon nanotubes.

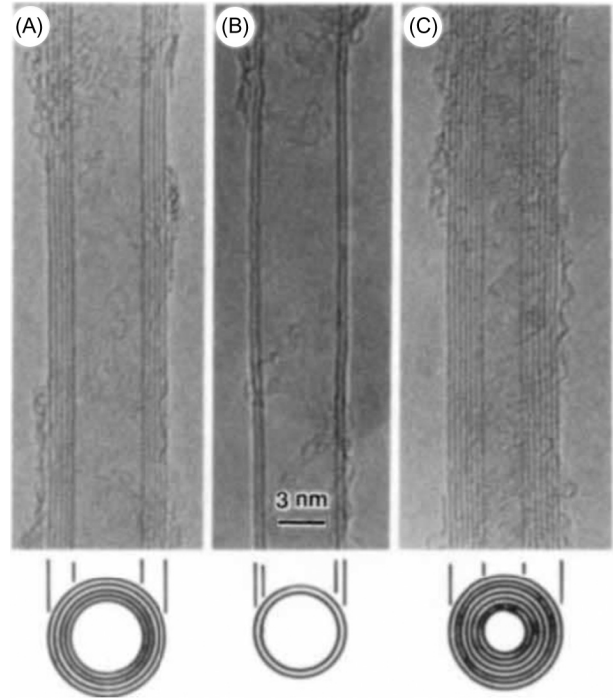

**FIGURE 9.3** Transmission electron microscopy images of CNTs synthesized by Iijima [10]. *CNTs*, carbon nanotubes.

### 9.1.4 Basic CVD Process

In CVD process initially all gaseous species are removed (called purging) from the reaction chamber other than those required for the deposition. Afterwards the precursor gases (hydrocarbon; denoted as CH) are delivered into the reaction chamber and heated (by energy source) as it approaches the deposition surface. When the precursor gas molecules acquire sufficient energy then they react or decompose to form a solid phase of carbon (C) which is readily deposited onto the substrate. Finally the volatile by-product hydrogen (H) is removed from the reaction chamber by an exhaust system (Fig. 9.4). In absence of substrate the solid phase of carbon is deposited on the walls of the reaction chamber.

## 9.2 OVERVIEW OF DIFFERENT CVD METHODS FOR INDUSTRIAL SCALE FABRICATION OF CNT

CVD methods can be categorized based on chamber pressure, reactor type, carbon source and heating method. A complete tree of different CVD categories is depicted in Table 9.1.

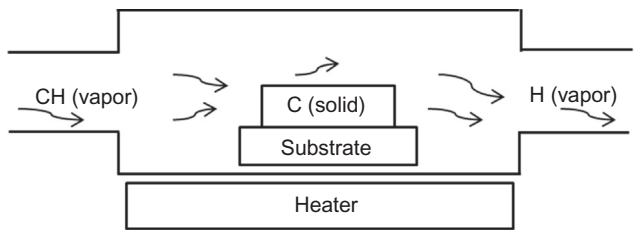

**FIGURE 9.4** Schematic representation of a CVD process. *CVD*, chemical vapor deposition.

### 9.2.1 CVD Techniques Based on Reactor Pressure

Pressure is an important parameter as far as CVD method is concerned. Based on CNT synthesized by CVD at different pressure, the methods can be categorized as high-pressure CVD (HPCVD), atmospheric-pressure CVD (APCVD), and low-pressure CVD (LPCVD).

#### 9.2.1.1 High-Pressure Chemical Vapor Deposition (HPCVD)

In this type of CVD the pressure inside the reactor is more than 1 atm during the synthesis of CNTs (Fig. 9.5). The high-pressure carbon monoxide (CO) decomposition technique (abbreviated as HIPCO) was developed for the mass production of single-walled CNTs (SWCNTs) by Smalley's group [11] at Rice University in the year 1999. According to the definition given by the Royal Society of Chemistry high-pressure carbon monoxide method is "A synthesis method for carbon nanotubes that involves mixing high pressure (e.g., 30 atmospheres) CO that has been preheated (1000°C) and a catalyst precursor gas (metal carbonyl or metallocene). Under these conditions the precursor decomposes forming metal particle clusters on which carbon nanotubes nucleate and grow. The carbon nanotubes are 99% single-walled carbon nanotubes with small diameters (e.g., (5,5) tubes)." In this process, high temperature exposure of the catalyst particle and the high pressure of CO result in rapid disproportionation of CO molecules into C atoms, thus forming SWCNTs in accelerated manner.

A similar high-pressure technique was adopted by Resasco et al. [13] for the production of SWCNTs by catalytic disproportionation of CO in the presence of a unique Co−Mo solid catalyst and the process is known as CoMoCAT. In CoMoCAT method the SWCNTs were synthesized by CO disproportionation at 700–950°C using a flow of pure CO in the presence of Co–Mo catalyst at a pressure of 1–10 atm. It is to be noted that a gaseous catalyst is used in the HIPCO process whereas a supported catalyst is used in the CoMoCAT process.

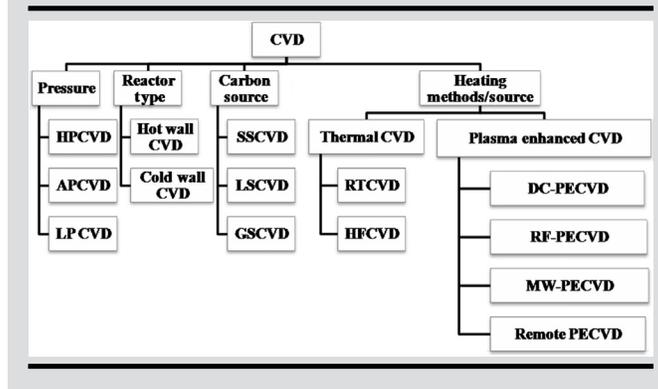

**TABLE 9.1** Types of CVD Techniques Employed for CNT Synthesis

## 9.2.2 Atmospheric-Pressure Chemical Vapor Deposition (APCVD)

In this synthesis method at atmospheric pressure the substrate is exposed to one or more volatile precursors which react or decompose on the surface of the substrate to produce a deposit (Fig. 9.6). In 1993, Yacamán et al. [14] reported the synthesis of carbon microtubules by catalytic decomposition of acetylene ($C_2H_2$) over Fe particles at 700°C. Aligned CNT growth was demonstrated in 2006 by Wei et al. [15] using Co as catalyst, $C_2H_2$ as carbon feedstock, and $NH_3$ as reaction control gas at 750°C. They concluded that $NH_3$ to $C_2H_2$ flow rate ratio and thickness of the catalyst film determines whether the CNTs would be vertically aligned or randomly oriented.

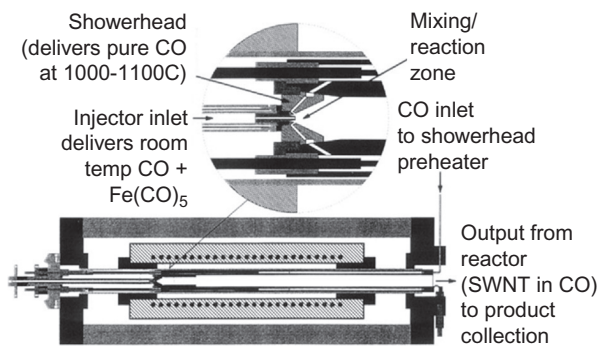

**FIGURE 9.5** Schematic representation of experimental setup (HIPCO reactor) used by Bronikowski et al. [12] for synthesis of CNT. *CNT*, carbon nanotube.

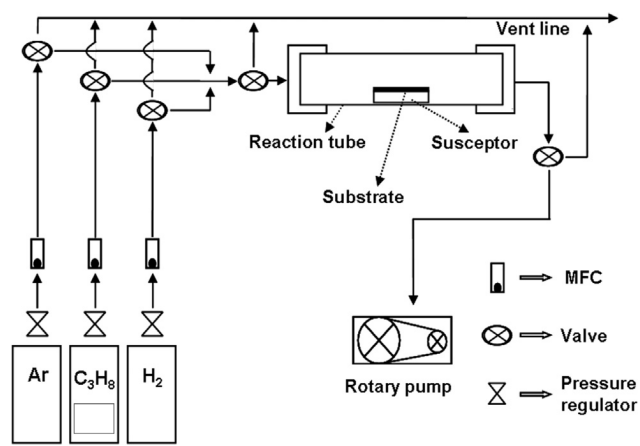

**FIGURE 9.6** Schematic representation of experimental setup used by Sengupta et al. [16] for synthesis of CNT. *CNT*, carbon nanotube.

Majewska et al. [17] exhibited a one-step method to synthesize CNTs filled with continuous Co nanowires in a conventional gas-flow system at the temperature of 400 and 800°C by methane ($CH_4$) decomposition at atmospheric pressure. Yun et al. [18] prepared highly aligned arrays of MWCNT on layered Si substrates using Fe as catalyst. They found that addition of water vapor to the reaction gas mixture of hydrogen ($H_2$) and ethylene ($C_2H_4$) accelerated the growth rate leading to the synthesis of longer CNT arrays with higher density. Cao et al. [19] used CVD to synthesize SWCNTs by employing a mixture of Fe/Mo/Co as catalyst, $Al_2O_3$ as substrate, and $C_2H_2$ as a carbon source. They demonstrated that the optimum growth occurs at a growth temperature of 750°C, $Ar/H_2/C_2H_2$ flow rates of 420/100/14 sccm, and a catalyst composition of Fe/Mo/Co/$Al_2O_3$ = 5/3/1/80 (wt%). Sengupta [20] and his coworkers also synthesized filled and unfilled MWCNTs by APCVD of propane ($C_3H_8$) on Si at 850°C using Ni- or Fe-coated Si substrate. They found that the nanotubes with magnetic material filling depict ferromagnetic behavior providing interesting possibilities for further applications in many potential areas, such as magnetic recording media. Patel et al. [21] prepared boron-filled hybrid CNT (BHCNT) using a one-step CVD process at 950°C under Ar and $CH_4$ flows of 100 and 10 sccm, respectively. Synthesized BHCNTs were up to 31% stiffer and 233% stronger than conventional MWCNTs in radial compression and exhibited excellent mechanical properties at elevated temperatures. Li et al. [22] demonstrated highly consistent synthesis of CNT forests at 775°C by decoupling the catalyst annealing and hydrocarbon exposures using movement of the sample in and out of the CVD system and stabilizing the moisture and hydrocarbon concentration between the two steps. Bi-layer catalyst Fe/$Al_2O_3$ was deposited using sputtering on oxide-coated Si wafers and $C_2H_4$ was used as source of carbon.

## 9.2.3 Low-Pressure Chemical Vapor Deposition (LPCVD)

In this type of CVD the pressure in reaction chamber is less than 1 atm (Fig. 9.7). Liao et al. [23] synthesized SWCNTs using decomposition of $C_2H_4$ at low pressure (10 mTorr) over Fe particles supported on Si wafers at temperature 550°C. Ikuno et al. [24] performed selective fabrication of straight CNT bridges between Fe nanoparticles by LPCVD using $C_2H_4$ at a low pressure of 100 Pa in the temperature range of 800–970°C. Resultant CNTs were composed of bundled SWCNTs and MWCNTs. Cantoro et al. [25] synthesized SWCNTs by CVD of undiluted $C_2H_2$ at 350°C under low pressure ($<10^{-2}$ mbar). Experimental results revealed that $NH_3$ or $H_2$ exposure promotes nanostructuring and activation of subnanometer Fe and Al/Fe/Al multilayer catalyst films before growth, resulting in SWCNT formation at lower temperatures. Chen et al. [26] demonstrated that using ferrocene $Fe(C_5H_5)_2$ as a catalyst precursor, cyclohexane ($C_6H_{12}$) as carbon source, $H_2$ as carrier gas, and thiophene ($C_4H_4S$) as promoter, the pressure of 15 kPa, temperature of 650°C and $H_2$ flow rate of 60 sccm is the optimized condition for the synthesis of high quality MWCNTs. Kasumov et al. [27] synthesized SWCNT at pressures down to 0.5mbar with $C_2H_2$ at temperatures 800–1000°C on Al/Fe-coated $Si_3N_4$ membranes supported on Si substrate, without any gas flow at any point of the entire process, including the heating and cooling of the samples. This procedure differs from other low pressure technique where the synthesis proceeds in the gas flow for 5–30 minutes. $C_2H_2$ is introduced in the sample chamber by single injection followed by fast evacuation.

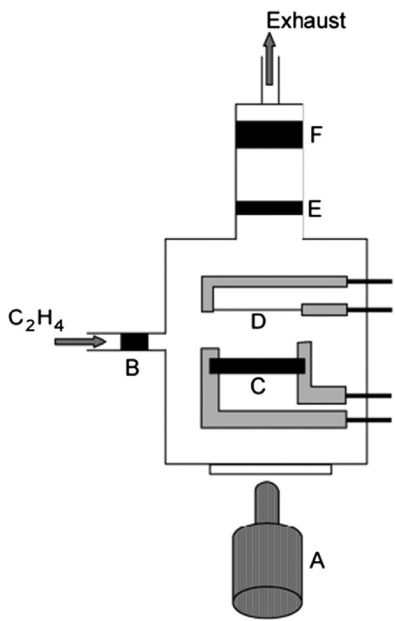

**FIGURE 9.7** Schematic representation of experimental setup used by Liao et al. [23] for synthesis of CNT. Legend is as follows: A, IR pyrometer; B, mass flow controller; C, Si substrate, D, Fe wire; E, throttle valve; and F, turbo pump.

High-density aligned CNT was synthesized by Wang et al. [28] on Fe-coated quartz substrate at 900°C at a pressure of 500 Torr. Their study showed that such high-density aligned nanotube has great potential of use for advanced nanoelectronics and analog/radio frequency (RF) applications.

### 9.2.4 Reactor Type

Based on temperature of the reactor wall, CNT synthesis by CVD process can be categorized as hot wall and cold wall CVD.

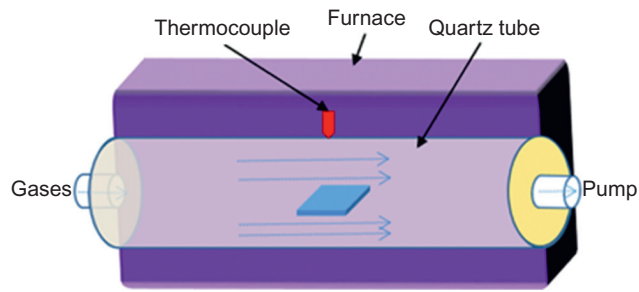

**FIGURE 9.8** Schematic illustration of the hot wall CVD furnace [31]. *CVD*, chemical vapor deposition.

#### 9.2.4.1 Hot Wall Chemical Vapor Deposition (Hot Wall CVD)

In hot wall CVD the reactor tube is surrounded by heating elements thus the temperature of the substrate is same as the reactor wall (Fig. 9.8). In general, for exothermic reactions the hot wall reactor is preferred as the high temperature of reactor wall restricts unwanted deposition on it. The great advantage of hot wall CVD is temperature uniformity. However as the reactor wall temperature is quite high thus vapor can chemically react with the reactor wall and create contamination in the grown product. Zhang et al. [29] used a hot wall reactor to carry out the CVD process to grow SWCNTs by decomposition of $CH_4$ on ultrathin Ni/Al film-coated Si substrates at temperature 800°C. High quality double wall CNTs (DWCNTs) with a defined diameter distribution were synthesized from alcohol at 940°C employing hot wall CVD by Grüneis et al. [30]. Catalysts for the DWCNT growth were made from Co and Mo acetates.

Ayala et al. [32] prepared high quality N-doped SWCNT and DWCNT with a defined diameter from a nondiluted C/N feedstock benzylamine ($C_7H_9N$) vapor using ceramic supported bimetallic catalysts containing Mo and Fe at temperatures 900–1000°C. A "super growth" of SWCNTs (Millimeter-thick forests of SWCNTs) was performed by Noda et al. [33] at temperature 820°C using $Fe/Al_2O_3$ catalysts on Si wafer employing $C_2H_4$ as carbon feedstock. Sugime et al. [34] investigated optimum catalytic and reaction conditions using a combinatorial catalyst library and also identified high catalytic activity areas on the substrate by mapping the CNT yield against the orthogonal gradient thickness profiles of Co and Mo. Yamada et al. [35] synthesized CNTs from 1.5 nm Fe thin film deposited onto 50 nm $SiO_2$ nanoballs placed on a transmission electron microscope (TEM) grid at 750°C using $C_2H_4$ as carbon source to study the effect of addition of water during the growth of nanotube.

### 9.2.4.2 Cold Wall Chemical Vapor Deposition (Cold Wall CVD)

In the cold wall CVD process only the substrate is heated by RF induction or high-radiation lamps while the reactor wall remains cold (Fig. 9.9). This type of CVD is primarily used for the endothermic deposition reaction. Compared to hot wall reactor this method requires shorter heating–cooling time and smaller growth periods and also prevents contamination of the chamber walls as walls are remain cold. Moreover, here only the sample needs to be heated, not the entire reactor. Thus, power and gas consumption is much lower than hot wall CVD process. Cold wall CVD is better suited to in situ optical monitoring than hot wall CVD, which may become an important advantage in the effort to better understand the growth process. However, temperature uniformity is the parameter of concern as there will always be cold surfaces that can conduct heat away thus creating a temperature gradient. Finnie's group [36] used the cold wall CVD technique to grow SWCNTs using $CH_4$ as source for carbon and Fe as catalyst on Si wafer at temperature 900°C. Later on they optimized the process [37] by studying the effects of pressure, temperature, substrate conditioning, and metallization. They also fabricated single-nanotube field effect transistors and reported the factors affecting device yield. Chiashi et al. [38] studied the cold wall CVD generation of high-purity SWCNTs from alcohol with Fe/Co particles supported on zeolite by Joule heating of a Si base plate at temperature 850°C. Cantoro et al. [25] reported surface-bound growth of SWCNTs at temperatures as low as 350°C by catalytic CVD from undiluted $C_2H_2$ using Fe-coated Si substrate. Maruyama et al. [39] studied SWCNT growth from Pt catalysts using a nozzle injector for the ethanol gas supply in a high vacuum. They grew SWCNTs at temperatures ranges from 330 to 700°C by optimizing the ethanol pressure and also demonstrated that the optimal ethanol pressure to obtain the highest SWCNT yield can be reduced if the growth temperature decreased.

### 9.2.5 Carbon Source

The catalyst is an important constituent for synthesis of CNT by CVD. Catalyst can be introduced into the reactor in various forms, e.g., solid, liquid, and gas. Based on the physical form of the catalyst used the CVD process can be categorized as solid source CVD (SSCVD), liquid source CVD (LSCVD), and gas source CVD (GSCVD).

### 9.2.5.1 Solid Source Chemical Vapor Deposition (SSCVD)

The term "SSCVD" encompasses all the techniques which employ only solid precursors as starting material for the deposition process (Fig. 9.10). In a dual zone tubular furnace the metallocene powder is kept in the first zone, called the preheating zone ($T_{pre} \ll T_{reac}$), and directly sublimated. The sublimation of powder forms metallocene vapor which is transferred by a controlled carrier gas flow into the second zone, i.e., reaction zone, where simultaneous decomposition of hydrocarbon and metallocene powder results in the growth of CNTs. Grobert et al. [40] described a way of generating films of aligned Fe-filled CNTs with enhanced magnetic coercivities in the 430–1070-Oe range. The material was synthesized by pyrolysis of $Fe(C_5H_5)_2/C_{60}$ mixtures at 900–1050°C under Ar flow using a conventional two-stage furnace. Müller et al. [41] synthesized Fe-filled aligned CNTs on oxidized Si substrates precoated with thin metal layers (Fe, Co) which act as secondary catalysts by thermal decomposition of ferrocene in an Ar flow, and discussed the CNT growth mechanism. Haase et al. [42] used

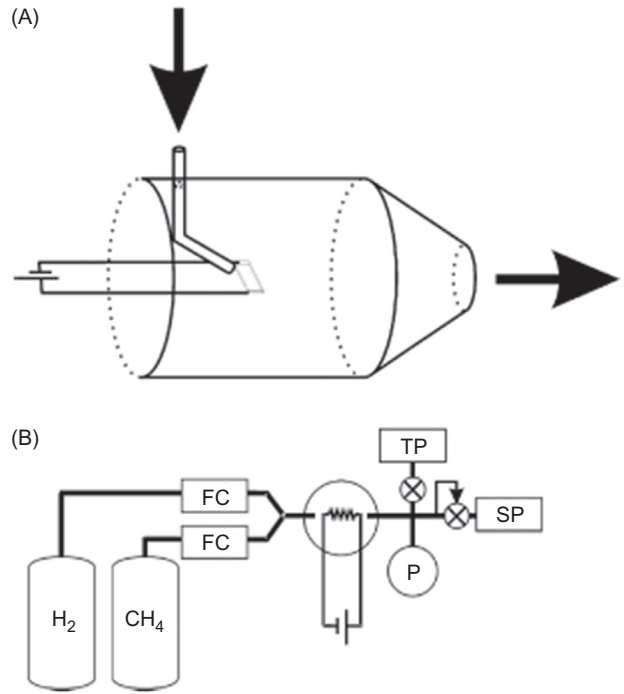

FIGURE 9.9 Cold wall CVD schematic: (A) reactor schematic (approximately to scale) and (B) gas flow schematic [37]. *CVD*, chemical vapor deposition.

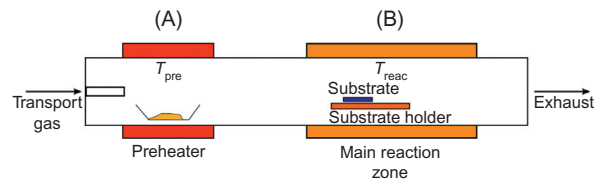

FIGURE 9.10 Schematic representation of SSCVD experimental setup used by Weissker et al. [44] for synthesis of CNT. *SSCVD*, solid source chemical vapor deposition; *CNT*, carbon nanotube.

SSCVD method to grow CNTs consisting of 20–30 walls with inner diameters of 10–20 nm with length varies from 10 to 30 μm. Later on the CNTs are filled with Carboplatin (a second generation cytostatic drug) to fabricate a CNT-supported drug delivery system of chemotherapeutic agents. Boi et al. [43] devised a new two-stage SSCVD approach, constituting a perturbed-vapor method of synthesis at 900°C using ferrocene, followed by a post-synthesis annealing at 500°C that produce MWCNTs filled with α-Fe nanowires.

### 9.2.5.2 Liquid Source Chemical Vapor Deposition (LSCVD)

In LSCVD only liquid precursor is used as primary material for the deposition process (Fig. 9.11). For the synthesis of CNT a hydrocarbon is used as a liquid precursor in which the metal catalyst compounds (e.g., metallocenes) are dissolved. Afterwards the precursor solution is introduced in the preheating zone of the dual zone CVD reactor having a temperature sufficient to atomatize the liquid precursor solution. Later on the atomatized precursor is transported to the reaction zone of the furnace using the flow of a suitable carrier gas. The simultaneous decomposition of both the hydrocarbon and the metallocene at the reaction zone produces CNTs.

Andrews et al. [45] synthesized high-purity aligned MWCNT via the catalytic decomposition of a $Fe(C_5H_5)$–$C_8H_{10}$ mixture over quartz substrates using Ar–$H_2$ mixture as transport gas at 675°C. Kamalakaran et al. [46] used the spray pyrolysis of $Fe(C_5H_5)$–benzene mixture to synthesize thick and crystalline nanotube arrays in an Ar atmosphere at 850°C, here $H_2$ was not required. Strong vertically aligned Fe-filled MWCNTs were synthesized using decomposition of $Fe(C_5H_5)_2$ in LSCVD process by Hampel et al. [47]. The filling yield was about 45 wt% and magnetometry measurements exhibited high magnetization moments, high coercivities, and strong magnetic anisotropies with an easy magnetic axis parallel to the aligned nanotubes. Peci et al. [48] devised a simple approach for production of continuous α-Fe nanowires encapsulated by MWCNTs of length greater than 10 mm through thermal decomposition of $Fe(C_5H_5)$. Nagata et al. [49] prepared Fe-filled CNTs with $Fe(C_5H_5)_2$ as a precursor on Si substrate at 785°C. The vertically oriented CNTs were almost completely filled with Fe. It was experimentally observed that coercivity of the Fe-filled CNTs can be enhanced by the addition of a Pt layer to the Fe catalyst film.

### 9.2.5.3 Gas Source Chemical Vapor Deposition (GSCVD)

The term "GSCVD" covers all strategies using only gaseous precursors as initial material for the deposition (Fig. 9.12). Here unlike SSCVD or LSCVD a single zone chamber is used and the gaseous precursor is directly introduced into the reaction chamber. The catalyst coated substrate is placed inside the chamber at high temperature sufficient to decompose the precursor in the presence of catalyst resulting in the growth of CNT. Zhao et al. [50] prepared horizontally aligned SWCNT arrays uniformly distributed all over the quartz substrate using nucleation of Cu nanoparticles on quartz as catalysts and $CH_4$ as carbon feedstock. They concluded that this kind of SWCNT arrays has great advantage in building large-scale integrated circuits. Ahmad et al. [51] used a combination of Fe, Ni, and Cr as catalyst for the synthesis of CNTs at different growth temperatures (600–750°C), keeping the ratio of $C_2H_2:N_2$ at 1:10 sccm. They concluded that the strength of epoxy resin improves with doping of well dispersed CNTs. Zheng et al. [52] synthesized SWCNTs directly on flat substrates using CVD method with CO and $H_2$ mixture as feeding gas. Monodispersed Fe/Mo nanoparticles were used as catalyst and $SiO_2/Si$, $Al_2O_3$, and MgO as substrates. The results showed that the formation of SWCNTs is greatly enhanced by addition of $H_2$. Kiribayashi et al. [53] studied the effects of fabrication method of $Al_2O_3$ buffer layer on Rh-catalyzed growth of SWCNTs by alcohol-gas-source CVD and found that the largest SWCNT yield could be achieved when $Al_2O_3$ layer is prepared by EB deposition of $Al_2O_3$, however no SWCNTs were grown on the $Al_2O_3$ layer obtained by native oxidation of the Al layer.

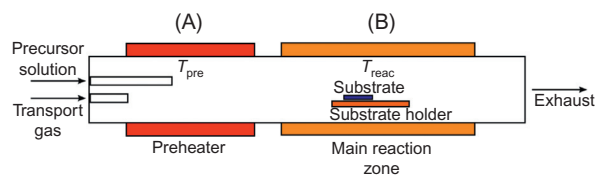

FIGURE 9.11 Schematic representation of LSCVD experimental setup used by Weissker et al. [44] for synthesis of CNT. *LSCVD*, liquid source chemical vapor deposition; *CNT*, carbon nanotube.

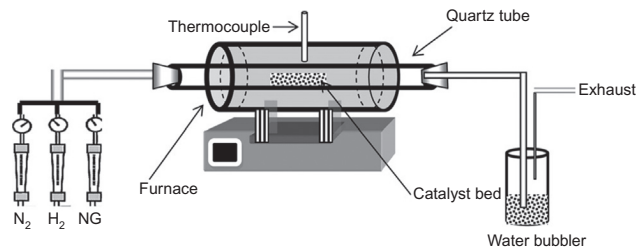

FIGURE 9.12 Schematic representation of GSCVD experimental setup used by Awadallah et al. [54] for synthesis of CNT. *GSCVD*, gas source chemical vapor deposition; *CNT*, carbon nanotube.

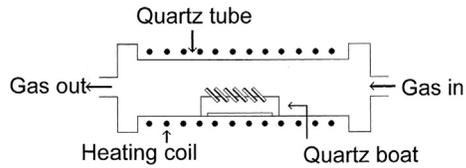

**FIGURE 9.13** Schematic diagram of thermal CVD apparatus used by Lee et al. [62] for synthesis of CNT. *CVD*, chemical vapor deposition; *CNT*, carbon nanotube.

### 9.2.6 Heating Methods/Source

When a traditional heat source such as resistive or inductive or infrared heater is used for deposition process, then the method is called thermal CVD (TCVD). If a plasma source is used to create a glow discharge then it is called plasma-enhanced CVD (PECVD).

#### 9.2.6.1 Thermal Chemical Vapor Deposition (TCVD)

In this synthesis method a substrate coated with catalyst is placed in a thermally heated atmosphere and exposed to one or more volatile precursors, which react or decompose on the surface of the substrate to produce a deposit (Fig. 9.13). For the synthesis of CNTs one or more hydrocarbon precursor are used and they decompose in the presence of one or more catalysts to produce CNTs. Lee et al. [55] synthesized aligned CNTs on transition metal-coated (Co–Ni alloy) Si substrates using $C_2H_2$ with a flow rate of 15–40 sccm for 10–20 min at the temperature of 800–900°C. They observed that pretreatment of Co–Ni alloy by HF dipping and etching with $NH_3$ gas prior to the synthesis was crucial for vertical alignment. Later on Lee et al. [56] used the same technique with Fe-coated Si substrate to grow aligned bamboo-shaped CNTs. Yoa et al. [57,58] studied the Si substrate/nanotube film interface in great detail and synthesized CNTs on them. CNT films were grown at 750 and 900°C by TCVD with $C_2H_2$ and $H_2$ on Si(0 0 2) wafers precoated with $(Fe, Si)_3O_4$ particles. At 750°C the reduction of the $(Fe,Si)_3O_4$ particles catalyzed the growth of a dense and aligned MWCNT film but CVD at 900°C a random growth of predominantly MWCNTs with lower density was obtained. Tripathi et al. [59] synthesized CNTs on $Al_2O_3$ substrate with diameter distribution 6–8 nm without using any catalyst by decomposition of acetylene ($C_2H_2$) at 800°C. Kozhuharova et al. [60] synthesized aligned Fe–Co alloy-filled MWCNTs using $Fe(C_5H_5)_2$/Co$(C_5H_5)_2$ mixture on oxidized Si substrates via TCVD at 980°C. The encapsulated metal nanowires had diameters of 10–20 nm and a length of up to a few micrometers. Reddy et al. [61] devised a single-step process for the synthesis of good-quality SWCNTs, MWCNTs, and metal-filled MWCNTs in large quantities by a TCVD technique, in which alloy hydride particles obtained from the hydrogen decrepitation technique was used as catalysts.

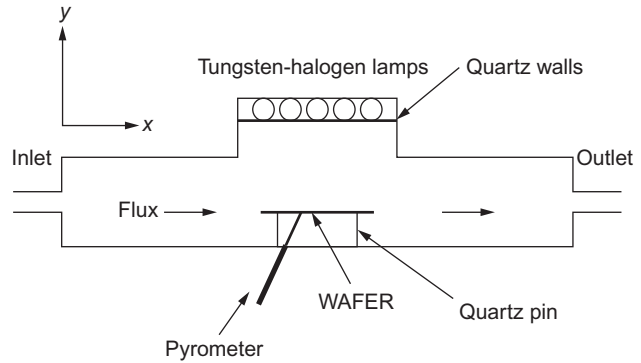

**FIGURE 9.14** Schematic diagram of the single wafer RTCVD process chamber [67]. *RTCVD*, rapid thermal chemical vapor deposition.

#### 9.2.6.2 Rapid Thermal Chemical Vapor Deposition (RTCVD)

It is a synthesis methodology where a rapidly heated substrate is exposed to one or more volatile precursors, which react and decompose on the surface to provide a deposit (Fig. 9.14). Here heating is achieved using infrared lamps, which makes the heating and cooling of the samples much faster. Typical duration for an RTP process is much less compared to a normal TCVD process which helps to minimize the thermal budget. When the heating lamps are energized then the irradiation (wavelength) passes through the quartz tube without being absorbed, while it is absorbed by the substrate. Thus, only the substrate heats up but its surroundings remain close to room temperature. Mo et al. [63] synthesized SWCNTs and MWCNTs on $Ni/Al_2O_3$ catalyst by thermal cracking of $C_2H_2/H_2$ at 600°C for 30 minutes with a gas flow rate of 10/100 sccm using a rapid TCVD (RTCVD). Based on the experimental observations they concluded that the grown CNTs followed the tip growth mode and the CNT growth was mainly due to the formation of a fluidized (liquid) metastable $Ni_xC$, i.e., metastable eutectic by dissolution of carbon, its oversaturation, and diffusion. Martin et al. [64] examined the relation between CVD process and CNT growth by placing the catalyst inside the pores of $AlPO_4$-5 and L-type zeolites using a RTCVD setup. CNTs were synthesized at 800°C under atmospheric pressure in the presence of $CH_4$ and $H_2$. They demonstrated that vacuum pretreatment and $H_2$ pretreatment of catalyst determine the growth morphology of carbon structures. Later on the same group [65] synthesized CNT at 800°C in a RTCVD system using $CH_4$ and $H_2$ as the main process gases and used them for batch fabrication of CNT field effect transistors. Chun et al. [66] synthesized MWCNTs by RTCVD using decomposition of $C_2H_2$ over a liquid catalyst solution (Fe–Mo/MgO/citric acid) at 700°C for 30 minutes. The thin MWCNTs showed the low turn-on field about 3.35 V/μm and the high emission current density of 1.0 mA/cm$^2$ at the biased electric field of 5.9 V/μm.

### 9.2.6.3 Hot Filament Chemical Vapor Deposition (HFCVD)

The HFCVD process employs a heated filament to decompose the precursor species and deposit a film on the substrate, which is placed close to the filament at lower temperature (Fig. 9.15). For the synthesis of CNTs a hydrocarbon is decomposed using a filament of refractory metal which is resistively heated at very high temperature. The type and quality of the synthesized CNTs are significantly influenced by the filament temperature. Chen et al. [68] synthesized well aligned bamboo shaped CNTs by HFCVD on Ni film-coated Si substrate using $C_2H_4/NH_3$ gas source with a flow rate of 25/100 sccm, while the total ambient pressure of the chamber was kept around 2.7 kPa. Diameter-controlled growth of SWCNTs was demonstrated by Kondo et al. [69] using Fe as catalyst and $C_2H_2$ as carbon feedstock at 590°C. Aligned CNTs growth on Inconel sheets was carried out using HFCVD in a gas mixture of $CH_4$ and $H_2$ by Yi et al [70]. The experimental results revealed degree of alignment of CNTs increased with the size of the catalyst particle and optimum alignment was achieved at a bias of −500 V. Choi et al. [71] studied hot filament effects on growth of vertically aligned CNTs (VACNTs) with respect to feedstock composition, filament temperature, and filament types. They used mixtures of methane and hydrogen as feedstock and found that growth rate increases with the increasing concentration of methane in the feedstock irrespective of filament temperatures and types. They also found that tungsten filaments were more efficient at the filament temperature of 2050°C for CNT growth than tantalum. Chaisitsak et al. [72] synthesized both SWCNT and MWCNT from $C_2H_2$ and $H_2$ mixture employing silica-supported Fe–Co as catalyst, by HFCVD method with a carbon filament. They found that formation of SWCNT is favored at low $C_2H_2$ concentration and low reaction pressure. Yilmaz et al. [73] demonstrated the growth of VACNTs in a planar configuration (48 μm tall) and a micropatterned array with 36-μm tall CNTs on Al substrates. They achieved the desired growth by combining conventional microfabrication steps for patterning an array and methane as carbon source for CNT synthesis.

### 9.2.6.4 Plasma-Enhanced Chemical Vapor Deposition (PECVD)

PECVD employs electrical energy to create glow discharge plasma. In glow discharge plasma the electron temperature is much higher than ion temperature which felicitates the maintenance of glow discharge plasma at low temperature. The high-energy electrons promotes the dissociation of gas molecules by which the energy is transferred into a gas mixture. This transforms the gas mixture into reactive radicals, ions, neutral atoms, and molecules, and other highly excited species. These atomic and molecular fragments interact with a substrate and, depending on the nature of these interactions, either etching or deposition processes occur at the substrate. Since high-energy electrons supply the energy needed for chemical reactions in the gas phase, the gas itself is relatively cooler. Hence, film formation can occur on substrates at a lower temperature than is possible in the conventional CVD process, which is a major advantage of PECVD. Moreover, in PECVD a high-electric field is generated in the sheath region due to potential difference between the plasma and the substrate.

PECVD reactors are mainly classified by the type of plasma source used to generate the gas discharge of the feedstock, i.e., direct current (DC), RF, and microwave (MW) PECVD.

### 9.2.6.5 Direct Current Plasma-Enhanced Chemical Vapor Deposition (DC-PECVD)

DC-PECVD employs DC power to generate glow discharge plasma between the anode and cathode, placed parallel to each other in a reactor (Fig. 9.16). A negative DC voltage is applied to the cathode to generate plasma and it also generates high-electric field in the sheath region between the substrate and the plasma. Tanemura et al. [75] used DC-PECVD to synthesize aligned CNTs on Co- or Ni-coated tungsten wires using mixtures of $C_2H_2$ and $NH_3$ and finally optimized the process with respect to wire temperature, wire diameter, gas pressure, and sample bias. They concluded that the selective feeding of positive ions to tip of CNTs is responsible for the alignment of growing CNTs. Kim et al. [76] prepared regular arrays of freestanding single CNTs using Ni dot arrays deposited on planar Si substrate at 550°C with a base pressure below $10^{-6}$ Torr. $C_2H_2$ and $NH_3$ were used as the carbon source and etchant gases. Abad et al. [77] synthesized CNTs both on bare stainless steel and on cobalt colloid nanoparticles coated stainless steel using $C_2H_2{:}NH_3$ gas mixture at 650°C. Ngo et al. [78] studied thermal interface properties of Cu-filled VACNT arrays prepared by DC-PECVD. Hofmann et al. [79] synthesized CNTs on Ni-coated $SiO_2$/Si substrate using a mixture of $C_2H_2$ and $NH_3$ and studied the effect of temperature on the growth rate and the structure of the CNTs. VACNTs were grown at temperatures as low as 120°C which promised the huge potential of the process to grow CNTs onto low-temperature substrates like plastics, and facilitate the integration in sensitive nanoelectronic devices. A hot filament suspended in the plasma is often integrated with the DC-PECVD system to grow CNTs [80–82]. This filament acts as a source of electron to stabilize plasma discharge and also as a heating source of the substrate. However, this approach restricts the possibility of low temperature growth and may act as a source of contamination. In general DC sources use most of the input power to accelerate the ions while ideally maximum power should be used to generate the reactive species in the bulk phase. Thus the high applied voltage in case of DC-PECVD may lead to substrate damage resulting from high energy ion bombardment.

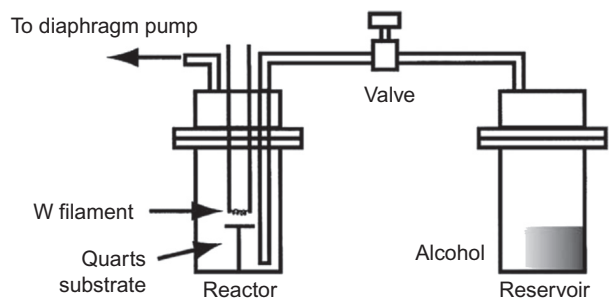

**FIGURE 9.15** Schematic diagram of the HFCVD apparatus used by Okazaki et al. [74] for synthesis of CNT. *HFCVD*, hot filament chemical vapor deposition; *CNT*, carbon nanotube.

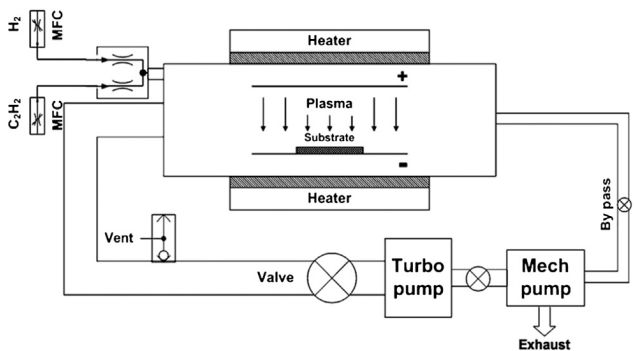
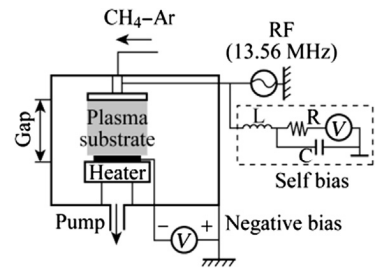

FIGURE 9.16 Schematic diagram of the DC-PECVD growth reactor used by Abdi et al. [83] for synthesis of CNT. *DC-PECVD*, direct current plasma-enhanced chemical vapor deposition; *CNT*, carbon nanotube.

FIGURE 9.17 Schematic diagram of the CCP-RF-PECVD system used by Man et al. [92] for synthesis of CNT. *CCP-RF-PECVD*, capacitively coupled plasma radio frequency plasma-enhanced chemical vapor deposition; *CNT*, carbon nanotube.

### 9.2.6.6 Radio Frequency Plasma-Enhanced Chemical Vapor Deposition (RF-PECVD)

RF-PECVD uses RF power to generate plasma [84]. RF (commonly 13.56 MHz) power is supplied to the reactor using an impedance matching network between the power supply and the plasma. The RF discharges remains operative at sub-Torr pressure levels and the bias voltage developed on the electrode is far smaller than the bias voltage of DC discharges. Depending on the method of coupling of the power supply to the plasma, RF plasma discharges are classified into two types (Figs. 9.17–9.19), namely, capacitively coupled plasma (CCP) [85] and inductively coupled plasma (ICP) [86]. Electrode configuration of CCP is similar to that of the DC-PECVD system however CCP RF-PECVD uses an RF power supply. The ICP system has a coil placed outside the reactor and RF power is supplied to the inside of the reactor from the coil through a dielectric window. Poche et al. [87] used CCP RF-PECVD to grow vertical field-aligned CNTs on Si wafer exhibiting a "herring-bone" and a "bamboo-like" structure at 560°C using Ni as a catalyst. Yuji et al. [88] used ICP RF-PECVD to synthesize bamboo-like CNTs with the arrowhead shape. Yen et al. [89] prepared aligned MWCNTs fully filled with Fe, Co, and Ni by ICP-CVD using nanowires as catalysts. A modified RF-PECVD system, namely magnetically enhanced RF-PECVD system [90], is also used for CNT synthesis. This system proficiently felicitates gas ionization in a magnetron-type PECVD reactor. Hirata et al. [91] synthesized vertically standing SWCNTs using this system by a mixture of methane ($CH_4$) and hydrogen ($H_2$).

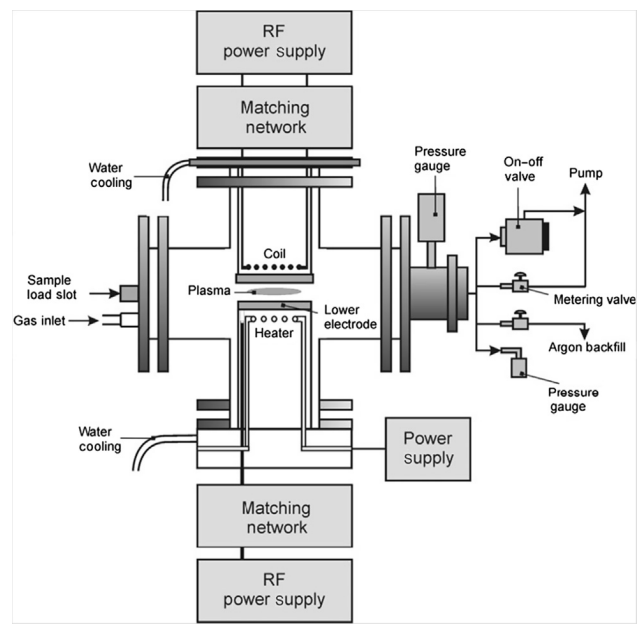

FIGURE 9.18 Schematic diagram of ICP-RF-PECVD reactor used by Meyyappan et al. [93] for synthesis of CNT. *ICP-RF-PECVD*, inductively coupled plasma radio frequency plasma-enhanced chemical vapor deposition; *CNT*, carbon nanotube.

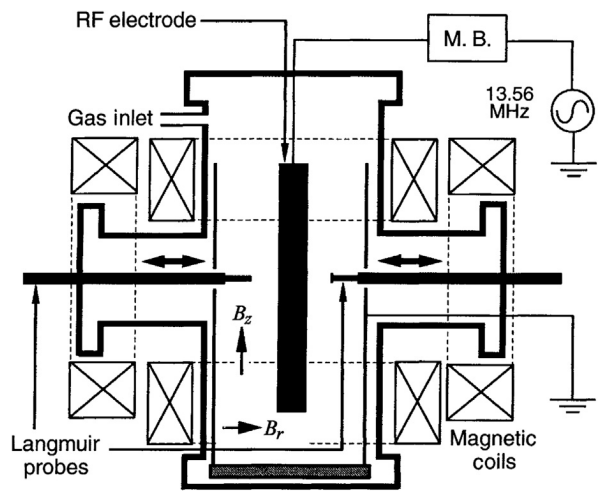

**FIGURE 9.19** Schematic diagram of magnetically enhanced RF-PECVD used by Ishida et al. [94] for synthesis of CNT. *RF-PECVD*, radio frequency plasma-enhanced chemical vapor deposition; *CNT*, carbon nanotube.

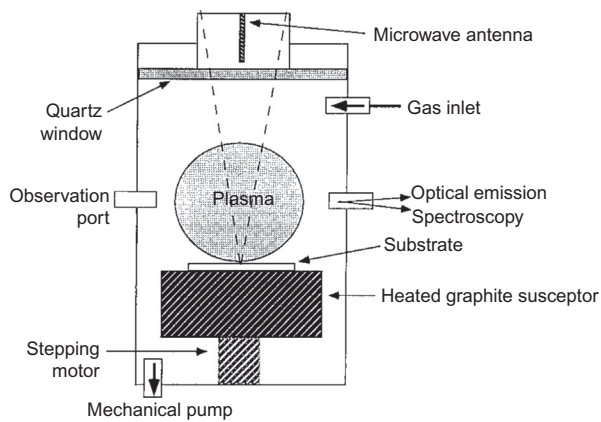

**FIGURE 9.20** Schematic representation of the MW-PECVD apparatus used by Qin et al. [102] for the synthesis of CNT. *MW-PECVD*, microwave plasma-enhanced chemical vapor deposition; *CNT*, carbon nanotube.

### 9.2.6.7 Microwave Plasma-Enhanced Chemical Vapor Deposition (MW-PECVD)

In MW-PECVD MWs (2.45 GHz) are introduced into the reactor through a dielectric window for irradiation of gases for generation of plasma (Fig. 9.20). MW-PECVD is capable to operate at a high pressure range and independent biasing of the substrate is possible using a DC or RF power supply. Tsai et al. [95] fabricated aligned CNTs with open ends on Si wafer in one step using MW-PECVD system with a mixture of $CH_4$ and $H_2$ as precursors. Wong et al. [96] synthesized well-aligned MWCNTs by MW-PECVD using $N_2$ as the carrier gas and $CH_4$ as the carbon source. Thin Fe films with different thicknesses (0.5–5 nm) on Si substrates acted as catalysts and grown CNTs showed a remarkable structural uniformity in terms of diameter. Hisada et al. [97] employed MW-PECVD to prepare single-crystal magnetic nanoparticles encapsulated in CNTs using Fe–Co catalytic layer on Si and $CH_4$ as carbon feedstock. Fujita et al. [98] produced fully encapsulated Co nanorods in MWCNTs by MW-PECVD using $CH_4$ on Si. Zhi et al. [99] prepared GaN nanowires encapsulated in CNTs using Fe-coated GaAs substrate by decomposition of $CH_4$. Hayashi et al. [100] have successfully grown well-aligned Pd/Co-filled CNTs with uniform diameter and length (1 m) on Si substrates at 750°C using $CH_4$ as carbon source. Zhang et al. [101] developed a method to grow Cu-filled CNTs using MW-PECVD. By adjusting the length of the exposed Cu electrodes they controlled the concentration of the Cu atomic clusters in plasma and then in the final products.

### 9.2.6.8 Remote Plasma-Enhanced Chemical Vapor Deposition (Remote PECVD)

It is possible to extract the plasma away from where it is generated, and ion-induced damage may be reduced when the wafer is located "remotely (Fig. 9.21)." One approach is to extract the plasma through a hole in the bottom electrode and place the wafer in a substrate holder further below [103]. Li et al. [104] used a remote RF plasma with either a monolayer of ferritin or 0.1 nm Fe layer as catalyst, and grew SWCNTs ranging between 0.8 and 1.5 nm in diameter. Min and coworkers [105,106] also used a remote PECVD system with a bimetallic CoFe (~9:1) layer having 0.9–2.7 nm in thickness. $SiO_2$ and $Al_2O_3$ layers were grown as diffusion barriers on glass or Si substrates. A water plasma with $CH_4$ gas, enabled a lower temperature of 450°C for growing SWCNTs which yielded dense nanotubes in the 1–2 nm diameter range. Fukuda et al. [107] succeeded in the CNT growth on the $Al_2O_3/Fe/Al_2O_3$ substrates on which no CNTs were grown by conventional TCVD. Ismagilov et al. [108] prepared layers of aligned MWCNTs on Si substrate without a metal catalyst by deposition from a $CH_4/H_2$ gas mixture activated by a DC discharge.

## 9.3 MATERIAL AND GROWTH CONTROL ASPECTS

### 9.3.1 Catalyst Type

The pathways for the synthesis of CNTs by CVD can be categorized into catalytic and noncatalytic methods

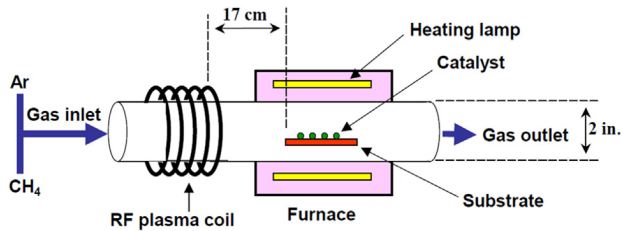

**FIGURE 9.21** A schematic diagram of the remote PECVD system used by Bae et al. [109] for synthesis of CNT. *PECVD*, plasma-enhanced chemical vapor deposition; *CNT*, carbon nanotube.

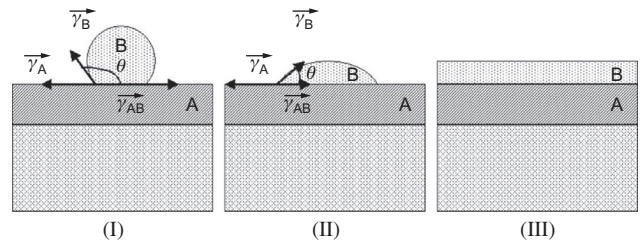

**FIGURE 9.22** Conditions of the surface energies of substrate (A), deposit (B) and interface A–B in determining the type of growth: island or Volmer–Weber growth for (I) nonwetting; (II) wetting; and (III) layer-growth [127].

[110–112]. In catalytic method the catalyst type can be broadly classified into two groups. If the catalyst particles are injected into the flowing feedstock to produce CNTs in the gas phase then the catalyst is called floating or unsupported catalyst [113]. In contrast, if the catalyst is deposited on the substrate for synthesis of CNTs before loading the substrate inside the reactor then it is called supported catalyst [114].

In the unsupported catalyst approach, a volatile compound containing the catalytic element [115,116], e.g., Fe(CO)$_5$, Fe(C$_5$H$_5$)$_2$, or Ni(C$_5$H$_5$)$_2$ is used as the catalyst source. The nanotubes form in the vapor phase and condense onto cold surfaces. However, the transition metal sources vaporize at temperatures much lower than that for the gas phase pyrolysis of the carbon sources and a two-zone furnace is generally required to produce CNTs by the unsupported catalyst approach. In this method, smaller catalyst clusters evaporates fast and usually unstable while large clusters are also detrimental as it favors graphitic overcoating. Thus it is the contest between several processes (clustering and evaporation) that produce favorable size clusters [117]. Sen et al. [118] first reported the unsupported catalyst approach and they used Fe(C$_5$H$_5$)$_2$ or nickolecene as a source of the transition metal and benzene as the carbon source. Cheung et al. [119] reported the synthesis of nanocluster solutions with distinct and nearly monodisperse diameters of 3.2, 9, and 12.6 nm for three different protective agents used, respectively. Addition of protective agents in the solution prevented the nanoparticles from aggregation. Hence, for large-scale continuous production of nanotubes, the floating catalyst approach is suitable. But the major drawback is that CNTs cannot be grown with site selectivity. Site selective growth of CNTs is a prerequisite for several device applications of CNTs.

The methods for the preparation of catalyst in case of supported catalyst approach can be divided into two categories: one is solution-based catalyst preparation technique and the other is the physical evaporation technique. There are numerous methods for preparing catalysts from solutions, for example sol–gel process [120,121], core-duction of precursors [122], impregnation [123], reverse micelle method [124], spin coating of catalytic solution [125], etc. Cassell et al. [126] demonstrated a combinatorial optimization process to assess the quality of MWCNTs produced by CVD. They constructed catalyst composition libraries to understand the effect of catalyst precursor composition on yield, density, and quality of the nanotubes with minimal number of growth experiments.

In the physical deposition technique, metal can be evaporated or sputtered to be deposited on the substrate. The metal, if deposited at room temperature, will generally be amorphous and form a nearly smooth film on the surface of the substrate. Upon annealing, the equilibrium shape may be reached. To know this shape, the Young's equation describing a contact between two phases A and B and the ambient atmosphere has to be considered:

$$\gamma_A = \gamma_{AB} + \gamma_B \cdot \cos\theta \quad (9.1)$$

where $\gamma$ is the corresponding interface energies (Fig. 9.22).

As $\theta \to 0$, $\gamma_A \to \gamma_{AB} + \gamma_B$, in this case, the growth is expected to occur layer by layer. When $\theta \rangle 0$, $\gamma_A \langle \gamma_{AB} + \gamma_B$ and discrete three-dimensional island-like clusters form as shown in Fig. 9.22.

It was found that the catalyst must be deposited on the substrate in the form of particles instead of smooth, continuous films [93] as continuous catalyst layer is unfavorable for CNT growth. Thus, the property of island growth is required for obtaining nanoparticles on a substrate by physical deposition. After deposition, usually performed at room temperature, annealing allows the atoms to rearrange themselves and reach the energetically most favored configuration. This method has been widely used to obtain nanoparticles of catalyst material to grow CNT. For breaking up the thin film obtained after deposition, many authors reported the use of NH$_3$ [128] or H$_2$ [129] during annealing. The surface energies $\gamma_A$ and $\gamma_B$ used above are defined relative to the atmosphere gas and thus a change in the gas can dramatically change the final shape of B on A. Delzeit et al. [130] demonstrated that introduction of a metal underlayer (such as Al) can be used instead of any chemical pretreatment steps to create a rough surface. Physical deposition techniques, such as thermal evaporation [131], electron gun evaporation [132], pulsed laser deposition [133], ion-beam sputtering [134], and magnetron sputtering [135], were successfully used for catalyst preparation.

A wide variety of catalytic species can be used to produce CNTs in CVD growth, based on their hydrocarbon decomposition ability, carbon solubility, stability, morphology, etc. Different transition metals (e.g., Fe, Ni, Co) and their alloys (e.g., Fe−Mo, Cu−Co, Ni−Ti) have been extensively used to grow CNTs by CVD [136−139]. Noble metals [140] (e.g., Au, Ag), semiconductors [141], diamond [142], sapphire [143], and even $SiO_2$ nanoparticles [144] behave as catalyst for nanotube growth. Deck et al. [145] studied the catalytic growth of CNTs using a broad range of transition metal catalysts in great detail. They demonstrated that Fe, Co, and Ni were the only active catalysts for CNT growth while Cr, Mn, Zn, Cd, Ti, Zr, La, Cu, V, and Gd have no catalytic features. They concluded that the solubility of the carbon in the metals is primarily responsible for their catalytic activity. However, metals like Cu, Pt, Pd, Mn, Mo, Cr, Sn, Au, Mg, and Al have been used to grow CNTs as reported by Yuan et al. [146]. Furthermore, recently in 2015, Hu et al. [147] synthesized high density SWCNT arrays with ultrahigh density of 130 SWCNTs/$\mu$m by ethanol CVD using Trojan catalyst, named by analogy with the soldiers emerging from the Trojan horse in the Greek Story. These observations indicate that the catalyst-growth dynamics-feedstock relation needs to be much explore in order to achieve supreme control over the CNT synthesis.

### 9.3.2 Catalyst Concentration

Kumar et al. [148] varied the catalyst concentration in zeolite from 1 to 50 wt % and observed that after a threshold concentration (2.4 wt%) formation of SWCNTs start and above 5 wt% of catalyst concentration growth of MWCNT is favored. They achieved highest yield of MWCNTs with negligible metal contamination using a combined Fe + Co concentration of 40%. The authors concluded that proper selection of catalyst materials and their concentration is essential for selective growth of SWCNTs or MWCNTs. Bai et al. [149] studied the influence of $Fe(C_5H_5)_2$ concentration of on the growth rate of the aligned CNTs. They observed that initially with increasing $Fe(C_5H_5)_2$ concentrations the growth rate of the aligned CNTs also increased up to a certain value while the growth rate decreased with subsequent increases in the ferrocence concentrations. They concluded that initially as the ferrocene concentration increases the availability of catalyst particle is also increased favoring the growth of CNTs specially SWCNTs because of their (catalyst) small size. However at higher ferrocene concentration the catalyst particles agglomerate and form large Fe clusters having weaker catalytic effect. Cao et al. [150] used $Fe_3O_4$ nanoparticles with varying concentrations (0.01, 0.026, and 0.033 g/mL) as catalysts for the growth of VACNTs at atmospheric pressure via catalytic decomposition of acetylene ($C_2H_2$) and observed that CNTs obtained using 0.026 g/mL of $Fe_3O_4$ had the highest uniformity, density, and length. Sangwan et al. [151] prepared ferric nitrate solutions of ten different concentrations (varying from 2 to 133 $\mu$g/mL) and used them to synthesize CNTs using $CH_4$, $H_2$, and $C_2H_4$ on oxidized Si substrate. They observed that at low concentrations density of CNTs increases linearly with concentration, and becomes almost constant for concentrations higher than 66 $\mu$g/mL and concluded that this method can be used to produce CNTs densities over a broad range varying from 0.04 to 1.29 CNTs/$\mu m^2$.

### 9.3.3 Support

Various substrates like Si [152], SiC [153], graphite [154], quartz [155], silica [156], alumina [157], aluminum [158], magnesium oxide [159], calcium carbonate [160], zeolite [161], NaCl [162], metallic alloy [163], etc. are used in CVD for the growth of CNT. Chai et al. [164] studied the effects of different support material such as silica, titania, ceria, magnesia, alumina, zeolite, and calcium oxide on CoO catalyst using $CH_4$ as a carbon feedstock. They found that $CoO/SiO_2$ catalyst shows the best performance with respect to carbon capacity at 550°C. Qingwen et al. [165] used $SiO_2$, $ZrO_2$, $Al_2O_3$, CaO, and MgO as support material for the growth of CNT at 850°C using methane as carbon precursor and Fe, Mo, or Ni as catalyst. The experimental results revealed that MgO is the best candidate as support in terms of efficient and stable growth, scalable synthesis at low cost and easy removal of the support. Their result revealed that the growth and orientation of CNT is also influenced by the grain boundary, crystallographic orientation, or crystalline steps of the support material. Su et al. [166] synthesized SWCNT with preferred 2D orientations by CVD of methane and showed that locking orientations are separated by 90° on Si(1 0 0) and 60° on Si(1 1 1), reflecting the lattice symmetry of the substrates. In contrast, Maret et al. [167] synthesized CNTs on single crystal MgO(0 0 1) substrate at 900°C by CVD using a mixture of CO and hydrogen, with Co catalyst nanoparticles and the results revealed that CNTs have formed along (110) direction. In 2011 He et al. [168] fabricated MWCNTs in six orthogonal directions on spherical alumina microparticles by CVD of xylene-acetylene mixture and concluded that "cube-like" surface structure of the alumina particles is responsible for such growth. Ghosh et al. [169] synthesized semiconducting SWCNT with a narrow diameter distribution between 0.67 and 1.1 nm by CVD on Pt-coated chemically treated graphene layers at 700°C using ethanol gas.

Since SWCNT and graphene both have excellent electrical, mechanical, and thermal properties so this result can pave the way towards future nanoelectronics. An interesting study about the influence of substrate surface area on the growth of CNT from ferrocene was investigated by Osorio et al. [170] using alumina, silica, carbon black, and zirconia powder as substrate. They concluded that the chemical composition of the substrate does not play a relevant role in the synthesis of CNTs, while the surface area has an influence and as the surface area of the supports increases, the synthesis temperature of CNTs gets lowered. Chhowalla et al. [171] investigated the growth process conditions of VACNTs by DC-PECVD of $C_2H_2$-$NH_3$ mixture on Si using Ni as catalyst and indicated that a barrier layer such as $SiO_2$ is required to prevent silicide formation. However, Nihei et al. [172] established the reverse by synthesizing VACNTs directly on nickel silicide employing MW-PECVD of $CH_4$–$H_2$ mixture. Due to these contradicting results of substrates with catalyst materials widespread use of underlayers came into existence. The interaction between catalyst and underlayer materials are numerous and complex and the choice of underlayer is extremely critical for CNT synthesis. Silicon oxide [173], magnesium oxide [174], and aluminum oxide [175] are among the most popular underlayer materials for CNT growth.

### 9.3.4 Temperature and Pressure

Lee et al. [176] synthesized VACNTs on Fe-deposited silicon oxide substrates by TCVD of $C_2H_2$ gas at the temperature range 750–950°C. The results exhibited that the growth rate and the average diameter increases with increasing growth temperature while the density of CNTs decreases. Moreover, the relative amount of crystalline graphitic sheets increases progressively with the growth temperature and a higher degree of crystalline perfection can be achieved at 950°C. Li et al. [177] found that growth temperature strongly influence the structure and yield of CNTs synthesized by CVD. With the increase of temperature the structure of CNTs changes from completely hollow to bamboo-like structure at low gas pressure of 0.6 Torr. While all the CNTs have bamboo-like structure irrespective of temperature at high gas pressure. Moreover, diameter of CNTs consistently increases with temperature within the experimental conditions but, at low gas pressure CNT diameter increases by increasing the number of graphene layers of the walls, while at higher gas pressure, the CNT diameter gets larger by increasing both the number of grapheme layers and the inner diameters. The same group also studied the effect of pressure in detail [178] and observed that reactor pressure severely influence the CNT production yield within the pressure range 0.6 to 760 Torr. Experimental results depicted that bamboo-like CNT structure was grown at high pressure whereas CNTs with completely hollow cores were formed at low pressure. They also found that with the increase of the gas pressure both the density of the compartments and graphene layers constituting the diaphragm in the bamboo-structured CNTs were increased. Ganjipour et al. [179] synthesized CNTs on Si substrates using a mixture of ethylene ($C_2H_4$) and $H_2$ gases in the presence of a Ni catalyst at a temperature of 700°C and gas pressures ranging from 10 to 400 Torr using DC-PECVD. It was observed that diameter of the CNTs increased as the deposition pressure increases till 400 Torr and further increase in the pressure leads to plasma degradation, and the growth can be paralyzed. Singh et al. [180] synthesized aligned MWCNT on quartz substrates by injecting a solution of ferrocene in toluene in a temperature range 550–940°C. The experimental results revealed that within the temperature range 590–850°C aligned CNTs were formed, with a maximum yield at 760°C. The diameter and diameter distribution of the CNTs increased with increasing temperature but above 940°C diameter again decreases and alignment was lost. Escobar et al. [181] synthesized MWCNTs by CVD using Fe as catalyst, $SiO_2$ as support, and acetylene ($C_2H_2$) as carbon precursor. The investigation about dependence of structural and morphological properties of MWCNTs on the partial pressure of acetylene ($C_2H_2$) revealed that at low $C_2H_2$ concentration, MWCNTs were more regular and with a lower amount of amorphous carbon than those synthesized with a high concentration. Sengupta et al. [182] studied the effect of growth temperature (varied from 650 to 950°C) on the growth of partially Fe-filled MWCNTs from propane decomposition on Si substrate with evaporated Fe layer using APCVD technique. The experimental results showed that the growth temperature of 850°C was optimum in terms of quality of the partially Fe-filled MWCNTs. Later on the same group [183] varied the preheating temperature from 850 to 1000°C and found that sample preheated at 900°C was the best in terms of growth density as well as the degree of graphitization. Afterwards Sengupta et al. [184] extended their study to investigate the effect of growth temperature on the growth of partially Fe-filled MWCNTs from propane decomposition on spin-coated Fe layer over Si substrate using APCVD technique. They found that in case of spin-coated catalyst the growth starts at 550°C.

## 9.4 GROWTH MECHANISMS

TCVD relies on thermal decomposition of carbonaceous gas molecules. Baker et al. [185] proposed a widely accepted general growth process of CNTs by CVD in 1972 which is based on the vapor−liquid−solid (VLS) mechanism (hydrocarbon vapor → metal−carbon liquid

→ crystalline carbon solid). Absorption, saturation, and structure extrusion are the three basic steps of VLS mechanism. In the VLS mechanism, a metal with low melting point and capable of absorbing the desired gas species is generally used as a catalyst. Here sufficient thermal energy is applied which transforms the solid metal catalyst into a molten state and decomposition of the carbon feedstock starts. After the initiation of gas decomposition, the carbon atoms start to diffuse within the catalyst and form a metal−carbon solution. As more and more carbon elements are incorporated into the catalyst, the concentration of carbon exceeds the solubility limit of the catalyst particle and consequently, the metal−carbon solution become supersaturated. Afterwards, addition of more carbon species into the system creates carbon precipitates at the surface of the particle which promotes the formation of tubular carbon structure. At this stage two different mechanisms are possible (Fig. 9.23), i.e., the catalyst particle is situated either at the top or at the bottom of the CNT after the growth is terminated. Bellavitis et al. [187] discussed the reason behind the occurrence of two different growth modes. Strong adherence of the particle with the substrate surface leads to the base growth model, since the carbon precipitates from the top surface of the particle and the tube continues to grow with the particle attached to the substrate. Whereas, weak adherence of the particle with the substrate surface leads to the tip growth model, since carbon precipitation occurs at the bottom surface of the particle so the particle is lifted up with the growing tube and finally reaches the top end of the tube. Vinciguerra et al. [188] suggest that tip growth or root growth of CNTs depends solely on the strength of the interaction of the catalyst particle with the substrate. However, the nature of the driving force for carbon diffusion through the catalyst particles is a subject of debate. The associated reports indicated that the temperature [189] or concentration gradient [190] within the particle can act as the driving force.

The key step in temperature-driven carbon diffusion mechanism was believed to be the diffusion of carbon species through the particle from the exposed and hotter front surface on which the exothermic decomposition of hydrocarbons occurs, to the cooler rear surfaces on which carbon is precipitated (endothermic process) from the solid solution [191]. The cooler surfaces are generally in contact with the support face. There is considerable experimental evidence to support this mechanism [192]. However, a temperature-driven dissolution−precipitation mechanism cannot provide a rational explanation for the endothermic pyrolysis of some hydrocarbons, e.g., $CH_4$ decomposition.

Concentration-driven carbon diffusion mechanism involves a concentration gradient across the catalyst particle in contact with hydrocarbon on one side and with a graphitic precipitation on the other side. Carbon growth involves a fast gas phase reaction (decomposition of hydrocarbon), carbon atom dissolution in the metal, and carbon precipitation as graphitic structures at the opposite side of the catalyst particle.

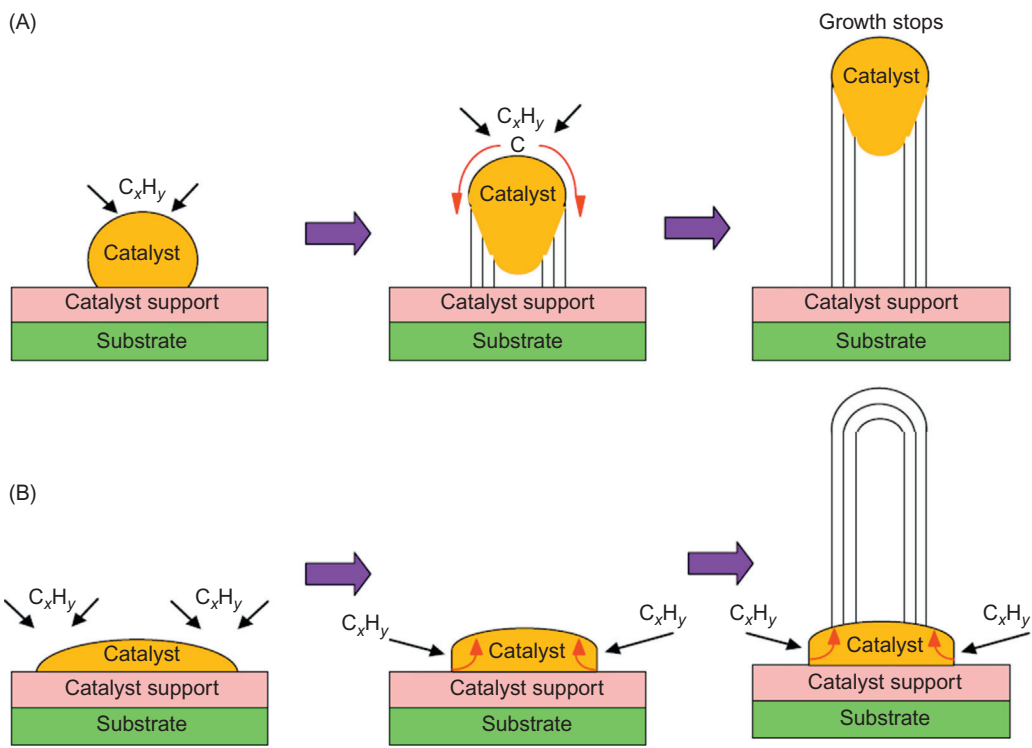

FIGURE 9.23 Schematic representation of the two typical CNT growth modes, (A) tip growth mode and (B) base growth mode [186]. *CNT*, carbon nanotube.

Primarily it was assumed that the volume diffusion of carbon through a catalyst particle is responsible for the catalytic growth of CNTs. However, Baird et al. [193] proposed a mechanism based on surface diffusion of carbon around the catalyst particle and was later expatiated by Oberlin [9]. Helveg et al. [194] studied the synthesis of CNTs using CVD of methane and Ni as catalyst under a controlled atmosphere TEM and their observation supports the surface diffusion mechanism (Fig. 9.24).

For a catalyst particle of unchanging size, the growth of CNTs should continue until the hydrocarbon is shut off, either by removing the feedstock from the reaction area or by amorphous or graphitic carbon fully coating the particle thus blocking the gas. Additionally, in base growth mechanism, prolonged diffusion of hydrocarbons down to the nanoparticle situated at the base of the CNT may slow down or stop the growth. Ideally, continuous growth of nanotube can be achieved if the nanoparticle is uninterruptedly fed with carbon source and the nanotube extrusion process is free of any obstruction. However, in reality, due to competing reactions at the nanoparticle site, such as the formation of graphitic shells and the deposition of amorphous carbon prevents the carbon feedstock from reacting with the nanoparticle resulting in termination of the growth.

## 9.5 KEY CHALLENGES AND THE FUTURE DIRECTION

1. **Synthesis and scaling:** Although many strategies have been opted for the synthesis of CNTs by CVD, efforts concerning the scale-up of CNT producing processes still face contemporary challenges like slow growth rate (e.g., low nucleation efficiencies), poor yield (e.g., low carbon utilization and fast catalyst deactivation), and undesirable variations in material quality and lack of real-time process control. The growth mechanism of CNTs in CVD is markedly difficult by the variability of experimental systems in terms of catalyst, support, carbon precursor, growth dynamics, and experimental conditions (humidity, etc.). Different systems usually result in too different observations to permit unification and rationalization. Further analysis on the mechanisms of CNT growth and producing scale-up would facilitate higher control of CNT properties like metallic or semiconducting [195,196], aspect ratio [197], chirality [198−200], alignment [201,202], length [203,204], diameter [205−207], defect density [208,209], and the number of walls [210,211]. Automated [212] and closed-loop method experimentation together with in situ [213−215] observation is essential for exploring the basic growth mechanisms and exploiting them for industrial production.

2. **Modeling and Simulation:** Advancement of the CNT synthesis process depends on enabling the growth infrastructure with proper simulation strategies and models that can be validated by experiment. In order to achieve the rapid growth of CNT-based products from the lab-prototypes, the scientists must develop a robust informatics infrastructure that provides the required optimization of the fabrication processes and helps to modify the final product performance. Numerical simulations are extremely beneficial for orienting experimental works and improving the understanding of CNT growth mechanism. With increasing calculative capacities, simulation has become an essential prediction tool for the phenomena that are still too difficult to study experimentally. For example, scientists used molecular dynamics simulations of CNT growth [216,217] and revealed that tube chirality controls the growth rate which is faster for armchair than zigzag. Moreover, analysis of CNT population dynamics during synthesis [218] from which absolute CNT mass and number density as well as CNT dimensions (diameter and number of walls) can be evaluated. Theoretical insight into the basics of the growth process specially nucleation efficiency and its mechanism [219,220] will also upgrade the development of processes to be capable of producing high-quality material in quantity. Interactions between individual tubes and the role of defects in tube−tube interactions are needed to be understood in a better way for advancement of CNT-based simulations. Moreover, a database which relates the catalyst composition and size to the properties of the grown CNTs is also required. More research is needed to understand the mechanism for the prolonged activation time for synthesizing longer CNTs. These cumulative efforts towards deep understanding of the growth mechanism will ultimately lead to a robust and validated model for large-scale growth reactors.

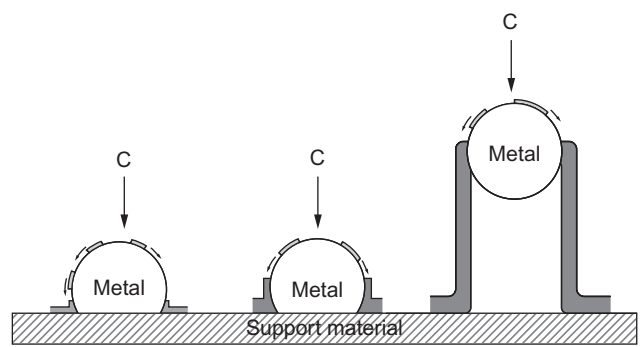

**FIGURE 9.24** Schematic representation of the CNT growth mode based on the surface diffusion of carbon around the metal particle [9]. *CNT*, carbon nanotube.

3. **Eco friendly:** Effect of CNT synthesis methods on environment [221] is a matter of concern considering the hazardous gas emission [222] from the reactor during the process. To comply with the environmental concerns the synthesis must be performed with renewable materials [223] to achieve environment friendly [224] green synthesis [225] with reduced energetic cost [226]. In order to protect the environment from a rapid degradation, industrial-scale production of CNTs should be carried out with sustainable technology and fossil fuel-based CNT production must be terminated. Scientists must identify and use cheap carbon sources [227,228], so that the price of CNTs could be scaled down to a reasonable rate. In order to save energy the lower growth temperature is an important factor. Until now scientists can go down to 120°C [76] in the case of low-pressure PECVD and 280°C in the case of atmospheric-pressure HFCVD [229]. Effort should be made in lowering the thermal budget in order to achieve long-term sustainability.

4. **Reproducibility:** Nicole Grobert [230] pointed out that "Carbon nanotubes will only have a significant impact when they can be produced with uniform properties." In order to use the exceptional properties of CNTs, they should be synthesized with new techniques where the defects can be monitored and removed in situ resulting in an atomically-perfect CNTs with uniform properties. Zhang et al. [231] has recently made moves in the required direction which must be scaled up towards higher-yield and lower-cost mass production of high-purity CNTs in a reproducible manner.

## 9.6 CONCLUSION

A US-based study [232] revealed that use of CNT in terrestrial and air transportation vehicles could enable a 25% reduction in their overall weight, reduce US oil consumption by nearly 6 million barrels per day by 2035, and reduce worldwide consumption of petroleum and other liquid fuels by 25%. This would result in the reduction of $CO_2$ emissions by as much as 3.75 billion metric tons per year. Thus it can be easily understood that CNT is an important member of today's nanotechnology regime. There are many pathways to synthesize CNTs but CVD is the best of the lot. Though there are several types of CVD, categorized based on some physical parameters, the main hurdle is to grow the CNT with desired structure with perfection (ideally zero defect density). The field has progressed in a rapid fashion towards scaling up to achieve mass production, however some more in-depth studies regarding growth mechanism are required to achieve the final step where perfect CNT with desired structure will be synthesized in an eco-friendly manner with reproducibility.